\documentclass[twocolumn]{aastex631} 
\usepackage{graphicx}
\usepackage{natbib}
\usepackage{placeins}
\usepackage{amsmath}
\usepackage{tgcursor}
\usepackage{booktabs}
\usepackage{enumitem}
\setcitestyle{notesep={ }}

\begin{document}

\title{HCN snowlines in protoplanetary disks: constraints from ice desorption experiments}

\author[0000-0002-8716-0482]{Jennifer B. Bergner}
\altaffiliation{NASA Sagan Fellow}
\affiliation{University of Chicago Department of the Geophysical Sciences, Chicago, IL 60637, USA}

\author{Mahesh Rajappan}
\affiliation{Center for Astrophysics \textbar\ Harvard \& Smithsonian, 60 Garden St., Cambridge, MA 02138, USA}

\author[0000-0001-8798-1347]{Karin I. \"Oberg} 
\affiliation{Center for Astrophysics \textbar\ Harvard \& Smithsonian, 60 Garden St., Cambridge, MA 02138, USA}

\begin{abstract}
\noindent HCN is among the most commonly detected molecules in star- and planet-forming regions.  It is of broad interest as a tracer of star-formation physics, a probe of nitrogen astrochemistry, and an ingredient in prebiotic chemical schemes.  Despite this, one of the most fundamental astrochemical properties of HCN remains poorly characterized: its thermal desorption behavior.  Here, we present a series of experiments to characterize the thermal desorption of HCN in astrophysically relevant conditions, with a focus on predicting the HCN sublimation fronts in protoplanetary disks.  We derive HCN-HCN and HCN-H$_2$O binding energies of 3207$\pm$197 K and 4192$\pm$68 K, which translate to disk midplane sublimation temperatures around 85 K and 103 K.  For a typical midplane temperature profile, HCN should only begin to sublimate $\sim$1--2 au exterior to the H$_2$O snow line.  Additionally, in H$_2$O-dominated mixtures (20:1 H$_2$O:HCN), we find that the majority of HCN remains trapped in the ice until H$_2$O crystallizes.  Thus, HCN may be retained in disk ices at almost all radii where H$_2$O-rich planetesimals form.  This implies that icy body impacts to planetary surfaces should commonly deliver this potential prebiotic ingredient.  A remaining unknown is the extent to which HCN is pure or mixed with H$_2$O in astrophysical ices, which impacts the HCN desorption behavior as well as the outcomes of ice-phase chemistry.  Pure HCN and HCN:H$_2$O mixtures exhibit distinct IR bands, raising the possibility that the James Webb Space Telescope will elucidate the mixing environment of HCN in star- and planet-forming regions and address these open questions.
\end{abstract}
\keywords{astrochemistry -- solid matter physics -- interstellar molecules -- protoplanetary disks}

\section{Introduction}
\label{sec:intro}

Following its initial interstellar detection by \citet{Snyder1971}, HCN has proven to be an important tracer of dense star-forming gas in both galactic and extragalactic environments \citep[e.g.][]{Kennicutt2012}.  As one of the few nitrogen carriers that are both observable and abundant, HCN is also a valuable probe of the nitrogen reservoir in star- and planet-forming environments \citep[e.g.][]{Rice2018, Cleeves2018}.  Beyond its utility as a probe of star-formation physics and chemistry, HCN has for decades been implicated in prebiotic chemical pathways \citep{Ferris1984}.  As a small, reactive organic, HCN is a key ingredient in astrobiological schemes including the synthesis of amino acids within interstellar ices \citep[e.g.][]{Elsila2007} and the production of nucleobases through planetesimal parent body processing \citep{Pearce2016, Paschek2021}.  HCN also plays a central role in prebiotic chemistry frameworks relevant to the early Earth surface \citep[e.g.][]{Ritson2012, Patel2015}.  While organics from the parent protoplanetary disk are not expected to survive the formation of primordial planetary atmospheres, later-stage impacts of icy bodies are a plausible source for this prebiotic feedstock \citep[e.g.][]{Pearce2017,Rubin2019,Todd2020}. The HCN inventory in protoplanetary disk ices is therefore highly relevant for understanding both the efficiency of ice-phase synthesis of biologically relevant molecules, and the prospects for exogenous delivery of HCN to planetary surfaces.

HCN is commonly detected in Solar system comets, with abundances $\sim$0.1--0.2\% with respect to H$_2$O \citep[e.g.][]{Mumma2011, Rubin2019b}.  Cometary HCN could be sourced either from the inheritance of HCN ice formed early in the star formation sequence, or through active synthesis during the protoplanetary disk stage.  HCN ice has yet to be detected in interstellar regions, though the upper limits are much higher than cometary ice abundances and therefore not particularly constraining \citep[$<$1.5\% with respect to H$_2$O;][]{Pontoppidan2019}.  Still, the gas-phase HCN/H$_2$O ratios measured in protostellar hot corinos appear somewhat lower than cometary \citep{Rice2018, Drozdovskaya2019}, suggestive of a disk contribution to the cometary HCN reservoir.  Indeed, protoplanetary disks appear to host a robust gas-phase photochemical production of HCN, particularly in somewhat elevated layers due to the high C/O ratio and strong UV field \citep{Du2015, Visser2018, LeGal2019, Bergner2019, Guzman2021}.  This material may diffuse to the midplane and be sequestered in ices, as is expected for other disk volatiles \citep[e.g.][]{Krijt2016,vanClepper2022}.  Thus, both interstellar inheritance and disk synthesis may contribute to the ice-phase HCN reservoir in disks.

A major limitation in understanding how HCN is incorporated into and retained in disk ices is that the sublimation behavior of HCN remains poorly characterized.  The only experimental binding energy for HCN in the literature is for a pure HCN ice \citep{Noble2013}.  Molecules desorbing from a H$_2$O ice surface or from within a H$_2$O ice matrix can show dramatically different behavior than pure ice desorption \citep[e.g.][]{Noble2012, Collings2004}.  To address this knowledge gap, here we present a series of experiments to characterize the desorption behavior of HCN in more astrophysically relevant conditions.  Section \ref{sec:exp} describes the experimental setup and procedures.  Section \ref{sec:res} describes our derivation of HCN-HCN and HCN-H$_2$O binding energies, as well as the entrapment efficiency of HCN in H$_2$O and a brief description of the IR band shapes during warm-up.  In Section \ref{sec:discussion} we discuss the implications of our findings, with a focus on the sublimation fronts of HCN in a planet-forming disk.  Section \ref{sec:concl} contains our conclusions.

\section{Experiments}
\label{sec:exp}

Experiments were performed on the ultra-high vacuum experiment SPACECAT\footnote{Surface Processing Apparatus for Chemical Experimentation to Constrain Astrophysical Theories}, which is described in detail in \citet{Lauck2015}.  The chamber achieves a base pressure of $\sim$5$\times$10$^{-10}$ Torr at room temperature.  A CsI substrate window is cooled by a closed-cycle He cryostat and can reach temperatures as low as $\sim$10 K.  A LakeShore 335 temperature controller is used to monitor the substrate temperature with an estimated accuracy of 2 K and a relative uncertainty of 0.1 K.  A differentially pumped gas line with a base pressure of 4$\times$10$^{-3}$ Torr is used to stage gases or gas mixtures prior to dosing.  Ice samples are grown on the CsI substrate by introducing the gases into the chamber through a dosing pipe with a 5 mm diameter.  Infrared absorbance spectra of the ice are obtained with a Bruker Vertex 70v Fourier transform IR spectrometer in transmission mode.  A Hiden IDP 300 (Model HAL 301 S/3) quadrupole mass spectrometer (QMS) is used to sample the gas-phase composition within the chamber.

\begin{deluxetable}{lcccc}
	\tabletypesize{\footnotesize}
	\tablecaption{Experiment summary \label{tab:exps}}
	\tablecolumns{5} 
	\tablewidth{\textwidth} 
	\tablehead{
        \colhead{Ice}       & 
        \colhead{Measurement} & 
        \colhead{Exp.} & 
        \colhead{HCN ML} &
        \colhead{H$_2$O ML} }
\startdata
Pure HCN & TPD & 1 & 32 & - \\
 &  & 2 & 7 & - \\
 & IR & 3 & 9 & - \\
\hline
HCN on c-H$_2$O & TPD & 4 & 0.142 & 37 \\
 & & 5 & 0.078 & 33 \\
 & & 6 & 0.042 & 31 \\
 & & 7 & 0.040 & 32 \\
 \hline
HCN:H$_2$O & TPD & 8 & 0.5 & 9.7 \\
mixture & & 9 & 2.2 & 53 \\
 & IR & 10 & 2.3 & 36 
\enddata
\end{deluxetable}

To prepare H$_2$O for our experiments, deionized H$_2$O was purified of hypervolatiles with several freeze-pump-thaw cycles in a liquid nitrogen bath (T$\sim$77 K).  HCN was synthesized in a purpose-built vacuum line with a base pressure of 2$\times$10$^{-3}$ Torr.  KCN crystals were dehydrated under vacuum, followed by the introduction of an excess of concentrated H$_2$SO$_4$.  The resulting HCN gas was collected in a liquid-nitrogen cooled flask.  This flask was attached to the SPACECAT corrosive gas mixing line, where the HCN was purified with several freeze-pump-thaw cycles using an ethanol:liquid nitrogen slurry (T$\sim$157 K). 

Table \ref{tab:exps} provides a summary of our experiments.  We explored the behavior of (i) multilayer HCN ice, (ii) submonolayer HCN on multilayer H$_2$O ice, and (iii) H$_2$O-dominated H$_2$O:HCN ice mixtures.  HCN was deposited at 40 K to minimize contamination of the ice by hypervolatiles (mainly N$_2$).  For the submonolayer experiments, we first deposited a thick H$_2$O ice at 100 K to produce a compact amorphous H$_2$O ice structure.  H$_2$O:HCN mixtures with an approximate 20:1 ratio were produced by mixing the gases in the gas line, then depositing the ice at 40 K.  We performed two types of measurements: temperature programmed desorptions (TPDs), in which the desorption rate of an ice species is monitored by QMS during a linear temperature ramp of 2 K min$^{-1}$; and IR spectroscopy, in which the infrared bands were continuously measured during a temperature ramp.

When possible, the surface coverage of each ice species is determined using their infrared bands, assuming band strengths of 1.03$\times$10$^{-17}$ cm molec$^{-1}$ for the 2100 cm$^{-1}$ HCN feature \citep{Gerakines2021} and 2.2$\times$10$^{-16}$ cm molec$^{-1}$ for the 3280 cm$^{-1}$ H$_2$O feature \citep{Gerakines1995, Bouilloud2015}.  We assume a typical uncertainty of 10\% on the IR band strengths, which is propagated through subsequent analysis steps.  We adopt the standard conversion of 1 monolayer (ML)$\equiv$10$^{15}$molecules cm$^{-2}$.  In our submonolayer experiments, the HCN feature could not be detected in the infrared spectrum.  Instead, we used a thick HCN ice to determine the scaling factor between the IR-derived ice thickness and the integrated QMS signal during a TPD.  We then scaled the integrated QMS TPD signal of the thin ices to recover the initial coverage.

\section{Results}
\label{sec:res}

\begin{figure}
\centering
    \includegraphics[width=\linewidth]{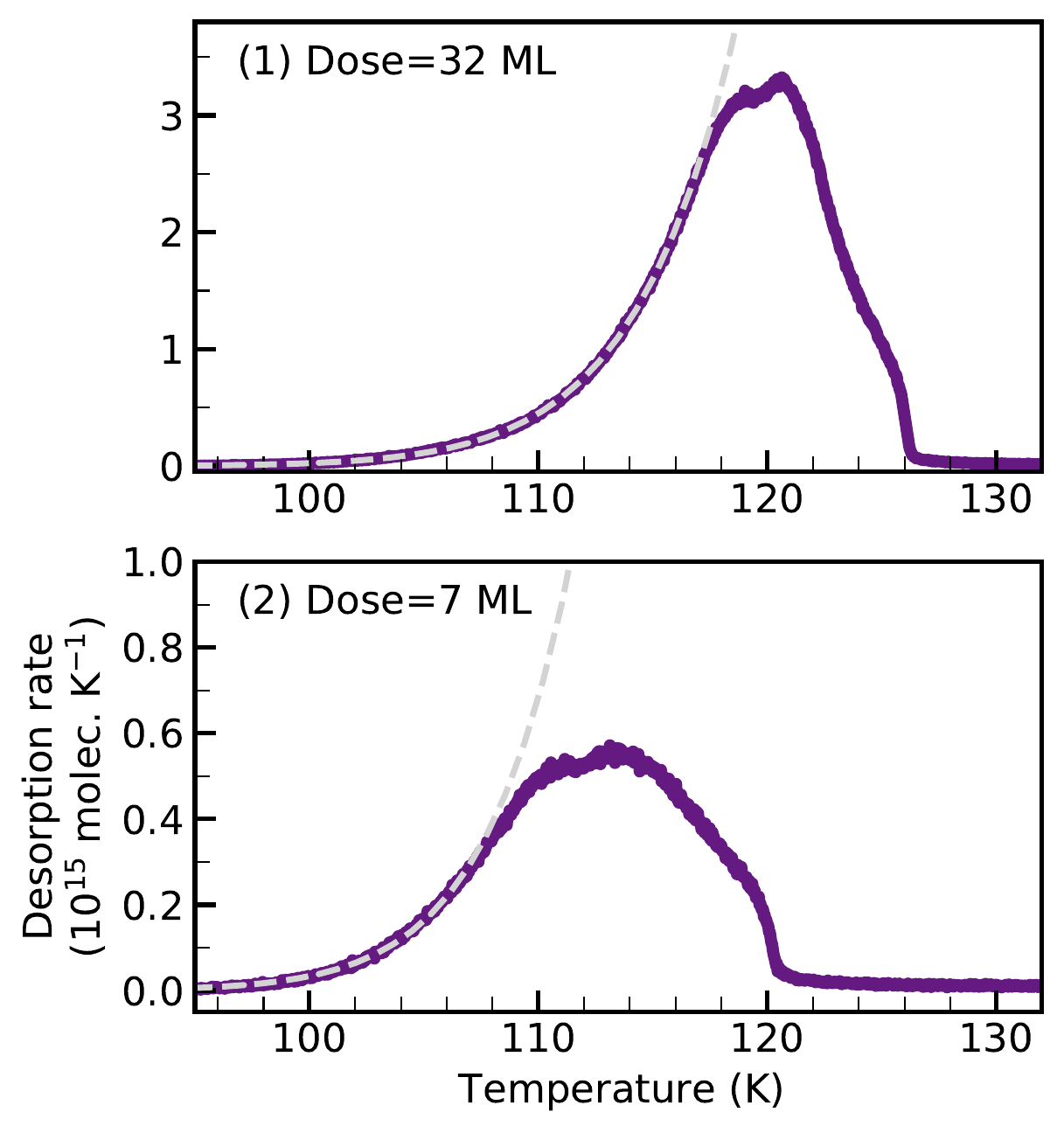}
    \caption{Multi-layer HCN TPDs (Exps.~1 \& 2).  The purple line shows the experimental data, and dashed grey curve shows the zeroth-order Polynani-Wigner fits.}
    \label{fig:hcn_multi}
\end{figure}

\subsection{Desorption energies}
\label{subsec:des}
We used our TPD experiments to constrain the desorption energies of HCN in the multilayer and submonolayer regimes, corresponding to the HCN-HCN and HCN-H$_2$O binding energies, respectively.  We fitted the experimental TPDs based on the Polyani-Wigner equation:
\begin{equation}
\label{eq:pw}
    -\frac{d\theta}{dT} = \frac{\nu}{\beta}\theta^ne^{-E_\mathrm{des}/T},
\end{equation}
where T is the ice temperature, $\theta$ is the ice coverage (in monolayers), $\nu$ is the pre-exponential factor associated with an attempt frequency, $\beta$ is the heating rate, and $E_\mathrm{des}$ is the desorption energy.  The kinetic order $n$ is 0 in the multi-layer regime, where the desorption rate is independent of the ice thickness; and 1 in the sub-monolayer regime, where the desorption rate is proportional to the ice coverage \citep[e.g.][]{Fraser2001}.  Note that the units of $\nu$ are dependent on the desorption order.  

\begin{figure*}
\centering
    \includegraphics[width=\linewidth]{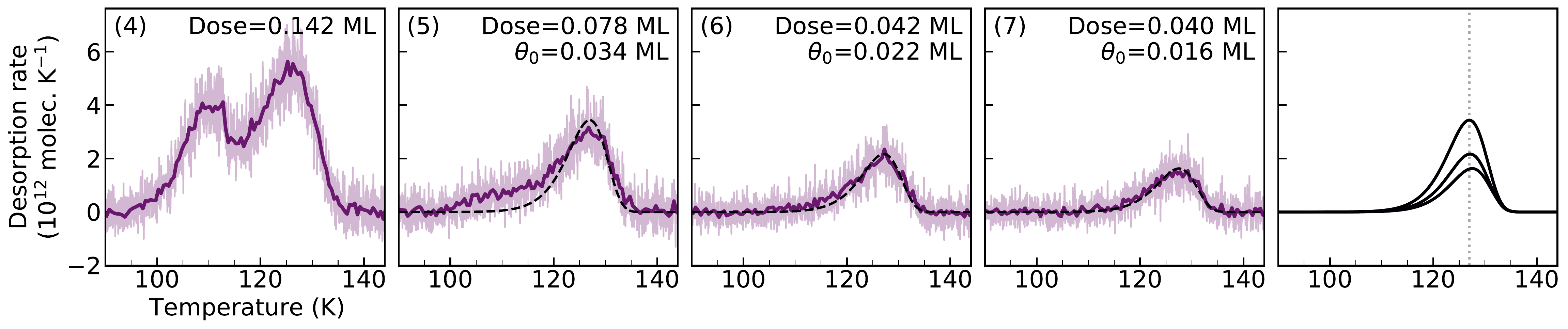}
    \caption{Sub-monolayer TPDs of HCN on compact H$_2$O ice (Exps.~4--7).  The light purple lines show the experimental data, with the mean value in 0.5 K bins shown in dark purple.  For experiments 5--7, the monolayer desorption feature could be reasonably well isolated, and dashed black curves shows the first-order Polynani-Wigner fits.  The fits are overlaid in the far-right panel.  The total HCN dose is listed for each experiment, along with the HCN coverage following first-order desorption kinetics.}
    \label{fig:hcn_subml}
\end{figure*}

The pre-exponential factor can be approximated as the harmonic oscillator vibrational frequency \citep[e.g.][]{Hasegawa1992, Acharyya2007, Noble2012, Fayolle2016}:

\begin{equation}
\label{eq:nu}
    \nu = \sqrt{\frac{2 N_s E_\mathrm{des}}{\pi^2 \mu m_H}}.
\end{equation}
Here, $N_s$ is the binding site density (10$^{15}$ cm$^{-2}$), and $\mu m_H$ is the molecular weight.  Note that this approximation is only valid for simple cases like atoms or small molecules.

In our analysis we directly fitted individual desorption curves to constrain E$_\mathrm{des}$ and $\nu$, thereby assuming that these parameters are independent of coverage.  As seen in Sections \ref{subsec:HCN0} and \ref{subsec:HCN1}, we obtain similar values for these parameters when fitting TPDs with different initial coverages, supporting that this is a reasonable assumption.  Still, TPD spectra for a larger number of initial coverages would be needed to perform a detailed assessment of the coverage dependence of E$_\mathrm{des}$ and $\nu$ \citep[e.g.][]{King1975, Falconer1983, deJong1990}.

\subsubsection{HCN-HCN binding energy}
\label{subsec:HCN0}
We solved for the HCN-HCN binding energy using the TPD spectra of multi-layer HCN ices with two different thicknesses (Figure \ref{fig:hcn_multi}).  We fitted the zeroth-order Polyani-Wigner equation (Equation \ref{eq:pw}) directly to each TPD curve in the regime where it is well-described by exponential behavior (T$\leq$116 K for Exp.~1 and T$\leq$107 K for Exp.~2).  We left both $\nu$ and $E_\mathrm{des}$ as free parameters, since fixing $\nu$ with Equation \ref{eq:nu} produced a poorer fit to the data.  The resulting fit parameter values are E$_\mathrm{des}$= 3207$\pm$115 K and $\nu$=6.8$\pm$3.5$\times$10$^{10}$ ML s$^{-1}$ for Exp.~1; and E$_\mathrm{des}$= 3318$\pm$128 K and $\nu$=2.9$\pm$1.9$\times$10$^{11}$ ML s$^{-1}$ for Exp.~2.  These uncertainties are dominated by the absolute uncertainty on the substrate temperature, since the fitting uncertainties are very small (a few K).  In Table \ref{tab:fits} we list the parameters derived from the thicker ice as the recommended values: compared to the thinner ice, the leading edge can be fitted over a wider temperature range and thus the exponential fit is better constrained.

\begin{deluxetable}{lcc}
	\tablecaption{HCN desorption parameters \label{tab:fits}}
	\tablecolumns{3} 
	\tablewidth{\textwidth} 
	\tablehead{
	    \colhead{Desorption order}       & 
        \colhead{$E_\mathrm{des}$  (K)} & 
        \colhead{$\nu$ $^a$} }
\startdata
0 (HCN-HCN) & 3207 $\pm$197 & 6.8 $\pm$4.9$ \times$10$^{10}$ \\
1 (HCN-H$_2$O) & 4192 $\pm$68 & 1.6 $\pm$0.01$ \times$10$^{12}$  \\
\enddata
\tablenotetext{}{$^a$Units are ML s$^{-1}$ for zeroth-order and s$^{-1}$ for first-order desorption}
\end{deluxetable}

For both ice thicknesses, we recover a similar set of E$_\mathrm{des}$ and $\nu$ values.  Still, the differences are larger than expected given the \textit{relative} uncertainty of the temperature controller, and are likely due to other sources of experiment-to-experiment variation.  Based on the comparison of Exps.~1 and 2, we assume that experimental variability contributes an additional 5\% uncertainty for E$_\mathrm{des}$ and 50\% uncertainty for $\nu$.  The recommended parameter values listed in Table \ref{tab:fits} account for this additional uncertainty.

It is important to note that the pure HCN ices do not follow zeroth-order kinetics throughout the entire desorption process, but have a complex behavior near the desorption peak and trailing edge.  This may be due to HCN crystallization during the temperature ramp, which occurs gradually between $\sim$10--120 K \citep{Noble2013}.  Amorphous-to-crystalline phase transitions produce similar deviations in the multi-layer desorption behavior of other small volatiles owing to the higher binding energy of the crystalline component \citep[e.g.][]{Fraser2001, Behmard2019}.  Still, because the leading edges of the pure HCN TPD curves are well-described by zeroth-order kinetics, we can fit this initial desorption component to derive the amorphous HCN-HCN binding energy.

\subsubsection{HCN-H$_2$O binding energy}
\label{subsec:HCN1}
To obtain the HCN-H$_2$O binding energy, we used TPD spectra of submonolayer HCN coverages on compact amorphous H$_2$O ice (Figure \ref{fig:hcn_subml}).  We found that HCN exhibits significant islanding, in which multi-layer clusters form as opposed to uniform wetting of the surface, even at very low coverages  \citep[e.g.][]{Noble2012}.  Indeed, we required a dose of 0.04 ML (Exp.~7) in order to fully isolate the monolayer desorption feature in the TPD spectrum.  At higher coverages (Exps.~4--6), an additional desorption feature at lower temperatures is apparent, corresponding to multi-layer desorption.  Note also that a third desorption feature starting at higher temperatures ($>$150 K) is present for all submonolayer TPDs, due to some entrapment in the underlying H$_2$O matrix.  While entrapment is discussed further in Section \ref{subsec:entrap}, here we do not consider the entrapped HCN except to note that the total HCN coverages listed in Table \ref{tab:exps} are higher than the effective coverages that desorb with first-order kinetics, due to both entrapment and islanding.

\begin{figure*}
\centering
    \includegraphics[width=\linewidth]{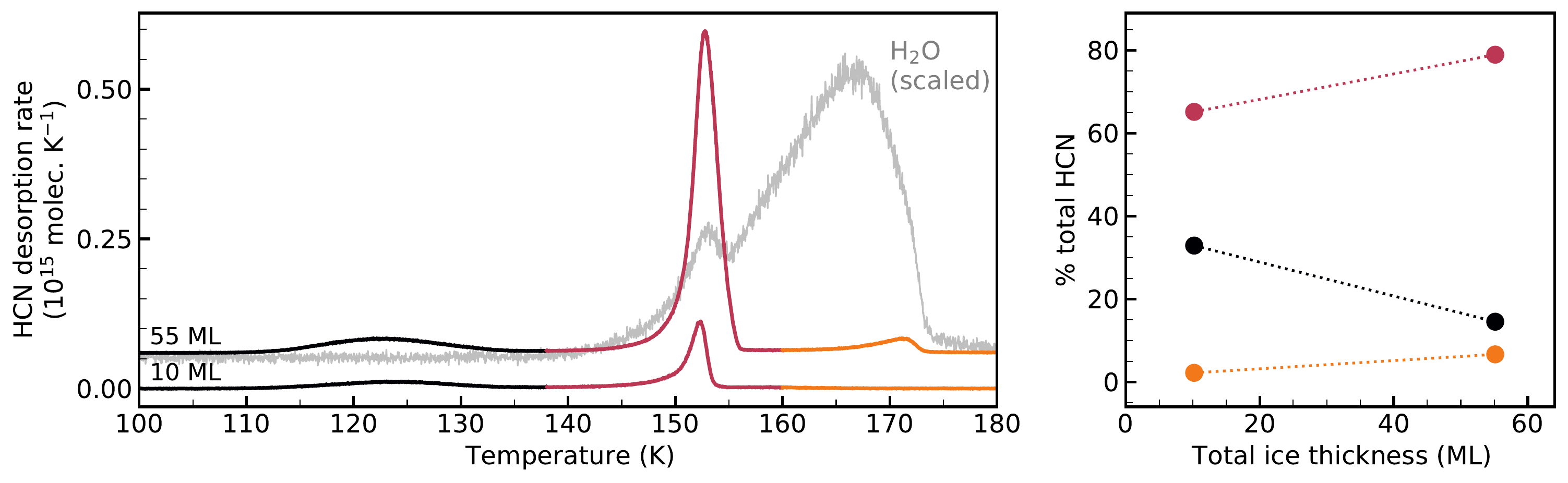}
    \caption{Left: HCN TPDs for $\sim$20:1 H$_2$O:HCN ice mixtures, with total thicknesses of 10 ML and 55 ML (Exps.~8 and 9).  Three desorption regimes are apparent: monolayer HCN desorption from H$_2$O (black), HCN `volcano' desorption due to H$_2$O crystallization (pink), and HCN co-desorption with H$_2$O (orange).  For comparison, a scaled H$_2$O desorption curve is shown in grey.  Right: fraction of HCN desorbing in each regime.}
    \label{fig:hcn_entrapment}
\end{figure*}

For Exps.~5--7, the monolayer desorption feature is well distinguished from the multilayer component, and could be fitted with the first-order Polyani-Wigner equation (Equation \ref{eq:pw}).  In this case, we fitted for the free parameters $E_\mathrm{des}$ and $\theta_0$, where $\theta_0$ is the effective HCN coverage that follows first-order desorption kinetics for each TPD.  Unlike for the multilayer TPDs, we obtained good fits with $\nu$ set to the harmonic oscillator frequency (Equation \ref{eq:nu}).  Note that due to the extremely low coverages, the submonolayer TPD spectra are fairly noisy.  While we fitted the data directly, Figure \ref{fig:hcn_subml} also shows the mean values in 0.5 K bins to enable a visual comparison with the fits.  The best-fit first order desorption energies are consistent among the three experiments (4173, 4178, and 4192 $\pm$68 K).  Experiment-to-experiment variability results in a small ($\sim$0.5\%) uncertainty on E$_\mathrm{des}$ compared to the absolute temperature uncertainty ($\sim$2\%).  Table \ref{tab:fits} lists the best-fit parameters based on the lowest-coverage TPD, since this fit is unaffected by the presence of a multilayer peak.  

Figure \ref{fig:hcn_subml} (right) shows the three best-fit submonolayer TPD curves overlaid.  The behavior is as expected for pure first-order desorption kinetics, with a common peak desorption temperature ($\sim$127 K) and misaligned leading and trailing edges \citep{Fraser2001}.  This is in contrast to numerous other small molecules like O$_2$, CO, N$_2$, and CH$_4$ which show an increasing peak desorption temperature with decreasing ice thickness \citep[e.g.][]{Noble2012, Fayolle2016, He2016}.  This is thought to reflect that an amorphous H$_2$O ice presents a range of binding environments: when the most energetically favorable binding spots are preferentially occupied, the peak desorption temperature is shifted to higher values at lower adsorbate coverages.  The fact that HCN does not exhibit this behavior may reflect that H$_2$O surfaces do not provide any especially favorable binding sites for HCN, which is also supported by its strong islanding behavior.

\subsection{HCN entrapment in H$_2$O}
\label{subsec:entrap}
We explored the entrapment behavior of HCN in a H$_2$O-dominated ice using $\sim$20:1 H$_2$O:HCN ice mixtures.  While this HCN concentration is much higher than what is expected in astrophysical or cometary ices ($\sim$500--1000:1), we chose this ratio to ensure that (i) HCN molecules should be completely surrounded by H$_2$O molecules within the mixture; and (ii) HCN is still detectable in the TPD and IR spectra when using reasonably thin ices (tens of monolayers).

Figure \ref{fig:hcn_entrapment} (left) shows the HCN TPDs for mixed ices with total thicknesses of 10 and 55 ML.  We identified three desorption regimes: a monolayer desorption feature between 110--130 K, a large desorption feature around 150 K, and a small desorption feature around 170 K.  This behavior is consistent with that of many other small astrophysical molecules \citep[e.g.][]{Collings2003, Collings2004}. The first peak can be understood as the loss of HCN occupying the H$_2$O surface or pores with access to the surface, resulting in desorption characterized by the HCN-H$_2$O binding energy.  The second peak, the so-called `molecular volcano' \citep{Smith1997}, occurs as a result of the transition from amorphous to crystalline H$_2$O.  During crystallization, new channels leading to the surface are opened, allowing HCN that was trapped within the amorphous ice structure to escape \citep{Collings2003}.  The third feature, which is apparent only for the thicker ice, represents co-desorption of HCN with H$_2$O.  Interestingly, the peak temperature of this feature is not coincident with the peak temperature of H$_2$O desorption, but occurs at the trailing end of H$_2$O desorption.

The entrapment efficiency of HCN in H$_2$O is quite high: 65--85\% of HCN remains trapped above its thermal sublimation temperature (Figure \ref{fig:hcn_entrapment}, right).  This is comparable to the range of trapping efficiencies measured for CO$_2$ in a H$_2$O matrix, and higher than the trapping efficiencies of CO in a H$_2$O or CO$_2$ matrix \citep{Fayolle2011, Simon2019}.  The vast majority of the trapped HCN desorbs during H$_2$O crystallization, with at most a few percent co-desorbing with H$_2$O.  The trapping efficiency is higher for the thicker ice, as expected if the occurrence of porous pathways connected to the surface decreases with increasing ice thickness \citep{Smith1997}.  

Our experiments include only one H$_2$O:HCN mixing ratio. In other experimental studies of volatile entrapment (using mixing ratios ranging from 1:1 to 10:1), the entrapment efficiency was found to increase as the entrapped species was further diluted \citep[e.g.][]{Fayolle2011, Simon2019}.  By extension, the entrapment of HCN in astrophysical ices ($>$500:1 H$_2$O:HCN) may be more efficient than in our experiments (20:1), if entrapment efficiencies continue to increase with dilution.  A more thorough exploration of HCN entrapment in different ice environments will be the subject of future work.

\begin{figure}
\centering
    \includegraphics[width=\linewidth]{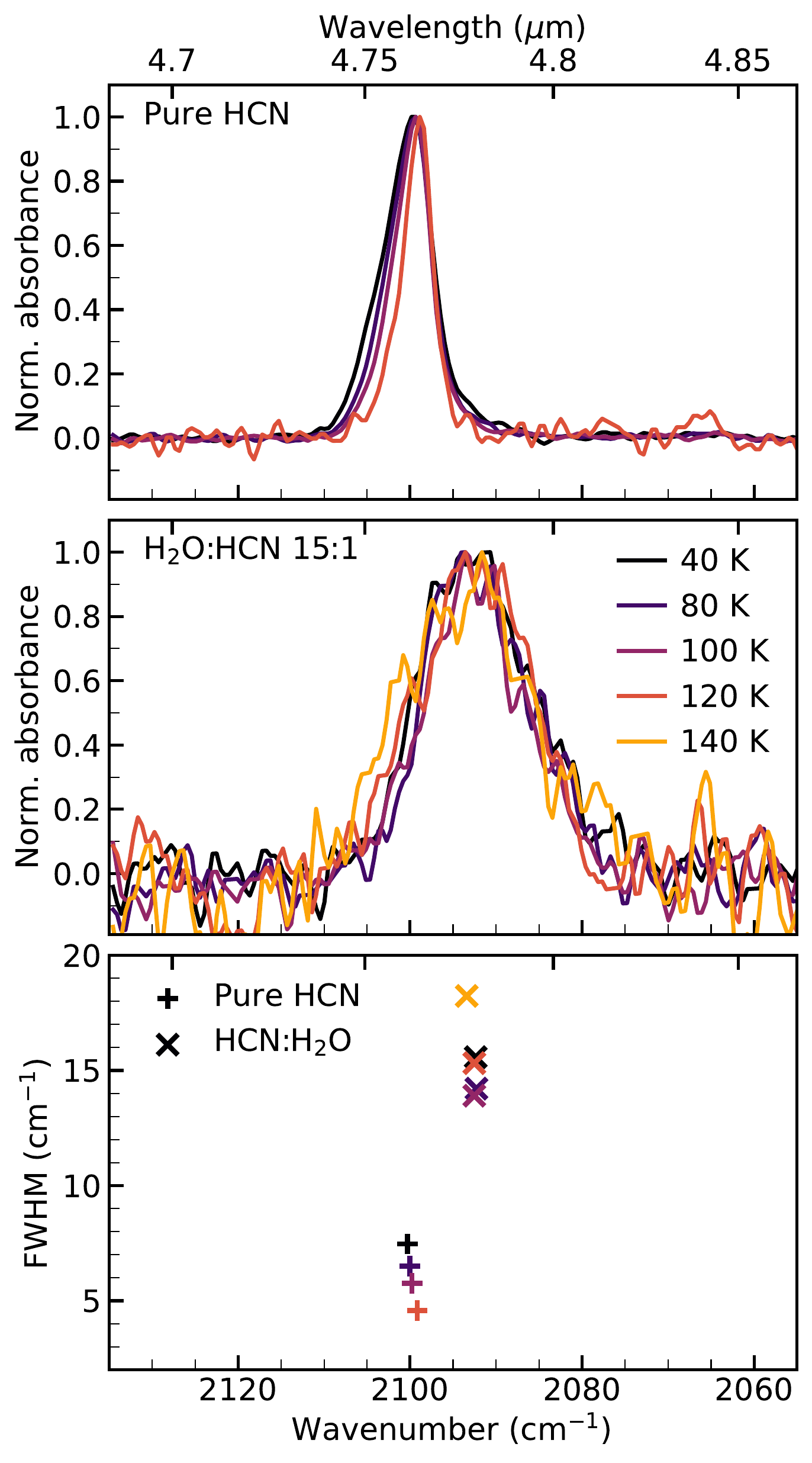}
    \caption{IR spectra of HCN (C$\equiv$N stretching mode) for a pure HCN ice (top panel) and a 15:1 H$_2$O:HCN ice mixture (middle pannel) at temperatures from 40--140 K.  Peak positions and widths for each spectrum are shown in the bottom panel.}
    \label{fig:hcn_spectra}
\end{figure}

\subsection{Infrared spectra}
\label{subsec:IR}
Since there are only limited descriptions available in the literature of the temperature evolution of the HCN infrared spectrum, we also provide IR spectra of pure HCN and HCN in a H$_2$O-dominated ice from 40--140 K.  Here, we focus on the C$\equiv$N stretching mode at 2100 cm$^{-1}$, but note that full spectra in the range 700-4000 cm$^{-1}$ (2.5--14 $\mu$m) in 10 K intervals are available on Zenodo \citep{zenodo}.  Figure \ref{fig:hcn_spectra} shows the pure and mixed HCN spectra, normalized to highlight changes in the band shape as a function of temperature.  As the temperature increases, the pure HCN band shows a slight but systematic narrowing along with a shift in peak position to longer wavelengths.  The HCN band shape in a H$_2$O-dominated mixture shows an even subtler response to the temperature.  In this case, the increasing temperature results in a slight broadening at shorter wavelengths, while the peak position remains almost constant.  At all temperatures, the HCN feature in a water matrix is broader and shifted to higher wavelengths compared to the feature in a pure HCN ice.

\section{Discussion}
\label{sec:discussion}

\subsection{HCN ice structure \& mixing}
\label{subsec:mixing}
We found that HCN exhibits significant islanding at low coverages (Section \ref{subsec:des}), requiring a dose of $<$0.04 ML to exhibit pure monolayer desorption kinetics.  Such behavior was predicted in a computational study of HCN desorption from a H$_2$O surface, which found that lateral HCN-HCN interactions are similar in strength to HCN-H$_2$O interactions, and play an important role in the adsorption of HCN \citep{Szori2014}.  Given the favorable energetics of HCN-HCN binding, pockets of pure HCN may be present within astrophysical ices if temperatures are sufficiently warm for ice diffusion.  Indeed, such `segregation' behavior is known to take place for CO$_2$:H$_2$O mixtures \citep[e.g.][]{Ehrenfreund1998, Oberg2009}, and CO$_2$ exhibits a similar islanding behavior at small coverages as is seen for HCN \citep{Noble2012, He2016}.  In our mixed HCN:H$_2$O experiments (Figure \ref{fig:hcn_entrapment}), the first HCN desorption feature is consistent with HCN-H$_2$O rather than HCN-HCN binding, indicating that segregation is not important during rapid heating of the ice.  Further experiments are needed to explore in more detail whether and in what conditions segregation takes place for HCN:H$_2$O ice mixtures.  

In addition to ice segregation, a pure HCN component could also originate during the protoplanetary disk stage, if HCN formed through gas-phase chemistry is efficiently accreted on top of the H$_2$O-dominated ice.  Ultimately, characterizing the mixing state of HCN in astrophysical ices is critical for predicting the extent to which HCN sublimation is characterized by the HCN-HCN vs.~HCN-H$_2$O binding energy.  As discussed further in Section \ref{subsec:sub_fronts}, this will control how HCN is retained in the ice phase across the formation zone of icy planetesimals in a disk midplane.  The mixing state of HCN will also impact the outcomes of ice chemistry: photolysis of pure HCN ices leads to HCN dimerization and potentially higher-order polymerization, while photolysis of HCN:H$_2$O mixtures may produce O-bearing species like OCN$^-$, HNCO, and NH$_2$CHO \citep{Gerakines2004}.  Therefore, proper modeling of ice phase chemistry requires constraints on the degree of mixing in astrophysical ices.

As seen in Figure \ref{fig:hcn_spectra}, the infrared band of pure HCN ice is distinguishable from the HCN:H$_2$O band.  Future ice spectroscopy observations may therefore permit some constraints on the HCN mixing status in interstellar environments.  Such observations will be possible with the James Webb Space Telescope (JWST), which will provide coverage of the 4.75--4.8$\mu$m HCN ice feature with unprecedented sensitivity and spectral resolution.  While HCN ice has yet to be detected in interstellar regions, the significantly improved sensitivity of JWST compared to previous IR telescopes raises the possibility of both detecting the HCN ice feature and characterizing its mixing status.  

\subsection{Comparison of HCN sublimation temperatures}
\label{subsec:disc_temps}
We derive an HCN-HCN binding energy of 3207$\pm$197 K (with $\nu$=6.8$\pm$4.9$\times$10$^{10}$ ML s$^{-1}$), which is in fair agreement with the value of 3608$\pm$120 K (with $\nu$=10$^{13}$ ML s$^{-1}$) measured in \citet{Noble2013}.  The slight discrepancy in binding energy could be explained by our use of the updated HCN band strength provided by \citet{Gerakines2021}, and/or uncertainties in the temperature controllers of the experimental setups.  When we fix $\nu$=10$^{13}$ ML s$^{-1}$ in our fitting, we obtain a worse fit to the data but a closer E$_\mathrm{des}$ of 3776 K.  In any case, the characteristic sublimation temperature resulting from the [$E_\mathrm{des}$, $\nu$] combinations in \citet{Noble2013} and our work are quite similar: following the formalism in \citet{Hollenbach2009}, with reasonable assumptions for disk midplane conditions ($n_\mathrm{H}$=10$^{10}$ cm$^{-3}$, H$_2$O/H=10$^{-4}$, HCN/H$_2$O=10$^{-2}$), both parameter sets result in a HCN-HCN sublimation temperature around 85 K.

The Kinetic Database for Astrochemistry\footnote{https://kida.astrochem-tools.org} provides a theoretically estimated HCN-H$_2$O binding energy of 3700$\pm$1100 K \citep{Wakelam2017}.  Our experimentally derived value of 4192$\pm$68 K is higher, but still within the large uncertainty range of the KiDA value.  With the same assumptions as above, the sublimation temperature of HCN from H$_2$O in midplane-like conditions is 90 K for the KiDA binding energy and 103 K for the value derived in this work.  Given this rather large discrepancy, we recommend the experimental value as both more accurate and more precise.  

For well-mixed ices, a large fraction of HCN may desorb at even higher temperatures due to entrapment in H$_2$O.  Indeed, in our mixed 20:1 H$_2$O:HCN ices only $\sim$15--35\% of the HCN desorbed in the monolayer regime, and the rest remained trapped in the H$_2$O matrix.  Thermal desorption of H$_2$O ice in a disk midplane should occur around 135 K \citep[$E_\mathrm{des}$=5600 K;][]{Wakelam2017}.  The `volcano' desorption of HCN will begin at somewhat lower temperatures, though the details will depend on the H$_2$O crystallization kinetics in a given disk location.  In any case, a significant HCN reservoir may be trapped within the H$_2$O matrix until tens of K higher than the thermal desorption of HCN.  It is important to note that on astrophysical timescales, diffusion could reduce the entrapment efficiency compared to lab experiments \citep{Cuppen2017}.  On the other hand, in thicker and more H$_2$O-dominated ices, entrapment efficiencies may be higher than those measured here. Additional constraints on the diffusion and entrapment behavior of HCN in a H$_2$O matrix are needed to explore these effects.

\begin{figure}
\centering
    \includegraphics[width=\linewidth]{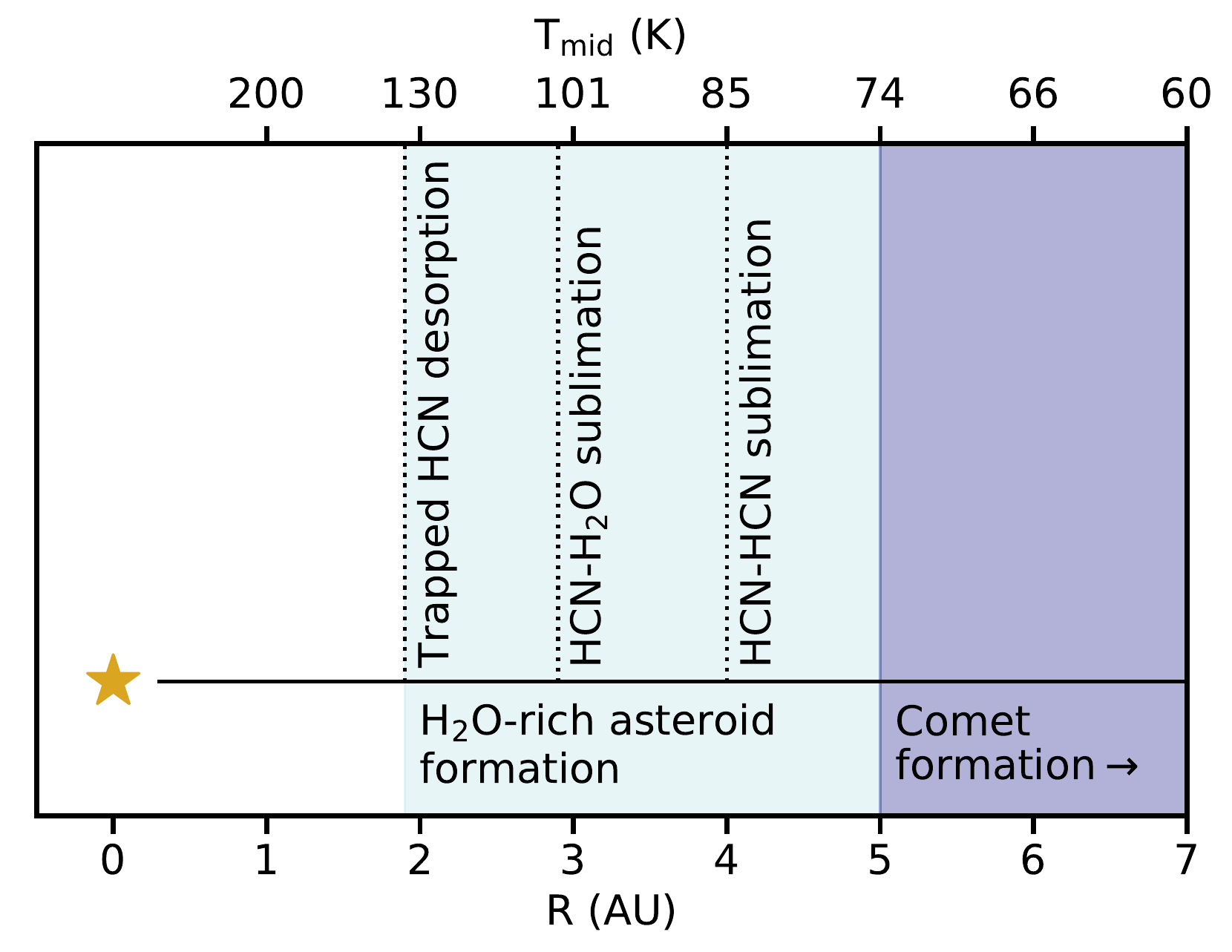}
    \caption{Cartoon of HCN desorption front locations in a typical T-Tauri disk midplane, compared to approximate icy body formation locations in the Solar Nebula.}
    \label{fig:snowlines}
\end{figure}

\subsection{Disk midplane sublimation fronts}
\label{subsec:sub_fronts}
The desorption behavior of HCN determines where in a disk HCN can be incorporated into icy planetesimals.  Here, we estimate the locations of HCN sublimation fronts for a typical T-Tauri disk midplane, assuming a temperature profile of 200 K$\times$(r/au)$^{-0.62}$ \citep{Behmard2019, Andrews2007}.  Figure \ref{fig:snowlines} shows a summary of the midplane desorption locations.  If pure HCN ice is present, this sublimation front will occur around 4.0 au.  Additional sublimation fronts corresponding to HCN sublimation from H$_2$O and desorption of entrapped HCN will occur around 2.9 au and 1.9 au, respectively.  HCN is therefore fully retained in the ice phase throughout the comet-formation zone \citep[$>$5 au;][]{Bockelee2004}.  Interior to this, i.e.~in the H$_2$O-rich asteroid formation zone, thermal desorption of HCN will be active.  In this region ($\sim$2--5 au), the degree of HCN desorption at a given radius will depend on (i) the extent to which HCN exists in a pure vs.~mixed ice environment (Section \ref{subsec:mixing}), and (ii) the efficiency of HCN entrapment on astrophysical timescales.  Given the high entrapment efficiency seen in our experiments, and the fact that ice mantles in disks are likely to be thicker and more H$_2$O-dominated than those studied here, we expect that complete loss of HCN from H$_2$O-rich ices is unlikely. 

HCN is also a weak acid, and can react to form the ammonium salt NH$_4^+$CN$^-$, which sublimates at a higher temperature than HCN.  For our assumed disk structure, the midplane sublimation radius of NH$_4^+$CN$^-$ is 2.7 au \citep[given E$_\mathrm{des}$=4570 K and $\nu$=10$^{13}$ ML s$^{-1}$;][]{Noble2013}.  This is comparable to the sublimation location of HCN from H$_2$O, and exterior to the sublimation front of entrapped HCN.  Thus, NH$_4^+$ salt formation will not preserve HCN in the solid phase interior to the H$_2$O snow line, as is expected for some other ammonium salts \citep[e.g.][]{Noble2014, Bergner2016, Kruczkiewicz2021}.  We also note that HCN has the highest sublimation temperature of the main interstellar ice constituents apart from H$_2$O and CH$_3$OH (e.g.~CO, N$_2$, CO$_2$, NH$_3$, CH$_4$).  This means that HCN will be an increasingly important ice-phase carrier of volatile C and N at decreasing disk radii.

\section{Conclusions}
\label{sec:concl}
 
HCN has long been implicated in prebiotic chemical schemes.  The delivery of HCN to planetary surfaces via the impact of icy bodies may contribute to the reservoir of material available for origins of life chemistry \citep{Pearce2017,Todd2020}.  Here, we performed a series of experiments to characterize the desorption behavior of HCN ice, which regulates where in the disk HCN may be incorporated into icy bodies.  

We find that HCN exhibits strong islanding behavior on a H$_2$O ice surface, which could translate to segregation of HCN within an astrophysical (H$_2$O-dominated) ice mantle.  Upcoming ice absorption spectroscopy with JWST may be able to distinguish pure HCN ice from HCN in a H$_2$O environment, and thus provide constraints on the HCN mixing status in star- and planet-forming regions.  This is ultimately important for predicting the importance of pure HCN desorption from astrophysical ices, as well as for robustly modeling the outcomes of ice-phase chemistry.

Based on our measured binding energies, in midplane-like conditions pure HCN sublimation should occur around 85 K, and HCN sublimation from a H$_2$O surface around 103 K.  The relatively high binding energy of HCN means that it should only begin to desorb $\sim$1--2 au exterior to the H$_2$O snow line.  We also find that $\sim$65--85\% of HCN within a 20:1 H$_2$O:HCN mixture is trapped in the ice until H$_2$O crystallizes.  While additional experiments are needed to quantify the impacts of ice thickness, mixing ratio, and diffusion on HCN entrapment, this high trapping efficiency implies that some HCN will remain in the ice until just exterior to the H$_2$O snow line.  Thus, HCN should be retained in disk ices at almost all radii where H$_2$O-rich icy bodies form.  This means that icy body impacts should commonly deliver this potential prebiotic ingredient.

While our work implies that impact delivery of HCN should be fairly \textit{common}, additional work is needed to understand the \textit{quantities} of HCN that will be delivered.  In addition to the HCN desorption efficiency at different radii, this will depend on the survival of HCN inherited from interstellar ices \citep[e.g.][]{Bergner2021} as well as the extent to which products of gas-phase disk chemistry are incorporated into midplane ices.  Ultimately, this is key to predicting whether impactors can provide HCN concentrations that are sufficiently high for prebiotic chemical schemes \citep[e.g.][]{Todd2020}.  

\begin{acknowledgements}
The authors thank the anonymous referees for feedback that improved the quality of this manuscript.  The authors are grateful to Zoe Todd, Edith Fayolle, and Elettra Piacentino for assistance with HCN synthesis.  J.B.B. acknowledges support from NASA through the NASA Hubble Fellowship grant \#HST-HF2-51429.001-A awarded by the Space Telescope Science Institute, which is operated by the Association of Universities for Research in Astronomy, Incorporated, under NASA contract NAS5-26555. 
K.I.\"O. acknowledges support from the Simons Foundation (SCOL \#321183).
\end{acknowledgements}

\software{
{\fontfamily{qcr}\selectfont Matplotlib} \citep{Hunter2007},
{\fontfamily{qcr}\selectfont NumPy} \citep{VanDerWalt2011},
{\fontfamily{qcr}\selectfont Scipy} \citep{SciPy2020},
}

\bibliography{references}

\end{document}